\documentstyle[12pt]{article}
\def\gappeq{\mathrel{\rlap {\raise.5ex\hbox{$>$}} {\lower.5ex\hbox{$\sim$}}}}
\def\lappeq{\mathrel{\rlap{\raise.5ex\hbox{$<$}} {\lower.5ex\hbox{$\sim$}}}}
\def\beq{\begin{equation}} \def\eeq{\end{equation}} \def\bea{\begin{eqnarray}}
\def\eea{\end{eqnarray}}
\def\bq{\begin{quote}} \def\eq{\end{quote}}
\def\ov{\overline}

\def\dd{\displaystyle}

\begin{document} 
\pagestyle{empty} 
\begin{flushright} {CERN-TH/99-147} \\ {DFPD 99/TH/30} \end{flushright}
\vspace*{20mm}
\centerline{\normalsize\bf LARGE NEUTRINO MIXING FROM SMALL QUARK AND LEPTON MIXINGS}
\baselineskip=16pt
\vspace*{1.5cm}
\centerline{\footnotesize GUIDO ALTARELLI}
\baselineskip=13pt
\centerline{\footnotesize\it Theoretical Physics Division, CERN, CH - 1211 Geneva 23}
\baselineskip=12pt
\centerline{\footnotesize\it and}
\baselineskip=12pt
\centerline{\footnotesize\it Universit\`a di Roma Tre, Rome, Italy}
\centerline{\footnotesize E-mail: guido.altarelli@cern.ch}
\vspace*{0.8cm}
\centerline{\footnotesize FERRUCCIO FERUGLIO}
\vspace*{0.3cm}
\centerline{\footnotesize and}
\vspace*{0.3cm}
\centerline{\footnotesize ISABELLA MASINA}
\baselineskip=13pt
\centerline{\footnotesize\it Universit\`a di Padova and I.N.F.N., Sezione di Padova, Padua, Italy}
\centerline{\footnotesize E-mail: feruglio@pd.infn.it, masina@pd.infn.it}
\vspace*{1.5cm}
\centerline{\bf Abstract} 

We discuss the special class of models where nearly maximal neutrino mixing is produced through the see-saw
mechanism, starting from only nearly diagonal matrices for charged leptons, Dirac neutrinos and Majorana 
right-handed neutrinos.
\vspace*{5cm}  \noindent  
\noindent
\begin{flushleft} CERN-TH/99-147\\ DFPD 99/TH/30\\July 1999\end{flushleft} 
\vfill\eject 

\normalsize\baselineskip=15pt
\setcounter{page}{1} \pagestyle{plain}

\section{Introduction}
Recent data from Superkamiokande \cite{SK} have provided a more solid experimental basis for neutrino oscillations as an
explanation of the atmospheric neutrino anomaly. In addition, also the solar neutrino deficit, observed by several experiments
\cite{solexp},
is probably an indication of a different sort of neutrino oscillations. Neutrino oscillations imply neutrino masses.
The extreme smallness of neutrino masses in comparison with quark and charged lepton masses indicates a different nature of
the former, presumably linked to lepton-number violation and the Majorana nature of neutrinos. 
Thus neutrino masses provide a
window on the very large energy scale where lepton number is violated and on Grand Unified Theories (GUTs). 
Experimental facts on neutrino masses and mixings could give an important feedback on the problem of quark and charged
lepton masses, as all these masses are possibly related in GUTs. In particular the observation of a nearly maximal mixing
angle for $\nu_{\mu}\rightarrow \nu_{\tau}$ is particularly interesting. Perhaps also solar neutrinos may occur with large
mixing angle. At present solar neutrino mixings can be either large or very small, depending on which particular solution
will eventually be established by the data. Large mixings in the neutrino sector are very interesting because a first guess 
was in favour of small mixings, in analogy to what is observed for quarks. If confirmed, single or double maximal mixings
can provide an important hint on the mechanisms that generate neutrino masses. Many possible theoretical descriptions
of large neutrino mixing(s) have been discussed in the literature \cite{review}. In most models large mixings are already
present at the level of Dirac and/or Majorana matrices for neutrinos. Instead, here we discuss an interesting class of
models where large, possibly maximal, neutrino mixings are generated by the see-saw mechanism starting from nearly diagonal
Majorana and Dirac matrices for neutrinos and charged leptons, without fine tuning or stretching small parameters into
becoming large.

The experimental status of neutrino oscillations is still very preliminary. Thus,
in order to be able to proceed, the theorist has to make a number of assumptions on how the data will finally look when
the experimental situation will be completely clarified. Here we assume that only two distinct oscillation frequencies
exist, the largest being associated with atmospheric neutrinos and the smallest with solar neutrinos. We assume that the
hint of an additional frequency from the LSND experiment \cite{LSND}, not confirmed by the Karmen experiment 
\cite{Karmen} (but yet far from
being completely excluded), will disappear. Thus we avoid the introduction of new sterile neutrino species and can deal
with only the three known species of light neutrinos. We interpret the atmospheric neutrino oscillations as nearly maximal
$\nu_{\mu}\rightarrow \nu_{\tau}$ oscillations, in agreement with the Chooz results \cite{Chooz}. 
The solar-neutrino oscillations
correspond to the disappearance of $\nu_e$ into nearly equal fractions of $\nu_{\mu}$ and $\nu_{\tau}$. A priori 
we are open minded
about which  of the three most likely solutions for solar neutrino oscillations is adopted: the two MSW solutions with
small (SA) or large (LA) mixing angle, or the vacuum oscillation solution (VO).

Assuming only two frequencies, given by 
\beq\Delta_{sun}\propto m^2_2-m^2_1~~,~~~~~~~
\Delta_{atm}\propto m^2_3-m^2_{1,2}~~,
\label{fre}
\eeq there are three possible hierarchies of mass eigenvalues:
\bea
        {\rm A}& : & |m_3| >> |m_{2,1}| \nonumber\\
        {\rm B}& : & |m_1|\sim |m_2| >> |m_3| \nonumber\\
        {\rm C}& : & |m_1|\sim |m_2| \sim |m_3|
\label{abc}
\eea 
(in case A there is no prejudice on the $m_1$, $m_2$ relation). For B and C, different subcases are generated
according to the relative sign assignments for $m_{1,2,3}$. The configurations B and C imply a very precise near
degeneracy of squared masses. For example, the case C is the only
one that could in principle accommodate neutrinos as hot dark matter together with solar and atmospheric neutrino
oscillations. We think that it is not at all clear at the moment that a hot dark matter component is really needed
\cite{kra}, but this could be a reason in favour of the fully degenerate solution. However, we would need a relative
splitting
$|\Delta m/m|\sim
\Delta m^2_{atm}/2m^2\sim 10^{-3}$--$10^{-4}$ and a much smaller one for solar neutrinos, especially if explained by vacuum
oscillations:
$|\Delta m/m|\sim 10^{-10}$--$10^{-11}$. It is reasonable to assume that the Dirac neutrino matrix has a strongly hierarchical
structure, with a dominant third family eigenvalue, as is the case for up quarks, down quarks and charged leptons. We
consider it implausible that, starting from hierarchical Dirac matrices, we end up via the see-saw mechanism into a nearly
perfect degeneracy of squared masses. Thus models with degenerate neutrinos  could only be natural if the dominant
contributions directly arise from non-renormalizable operators of higher dimension. However, such a precise near degeneracy, 
even if true at the GUT scale, could only be stable enough against renormalization group
corrections when running down to low energies if it were protected by a symmetry \cite{ell}. 
\section{A $2\times 2$ Example}
As a consequence, here we concentrate on models of type A
with large effective light neutrino mass splittings and large mixings. In general large splittings correspond to small
mixings because normally only close-by states are strongly mixed. In a 2 by 2 matrix context, the requirement of large
splitting and large mixing leads to a condition of vanishing determinant. For example the matrix
\beq 
m\propto 
\left[\matrix{ x^2&x\cr x&1    } 
\right]
\label{md0}
\eeq has eigenvalues 0 and $1+x^2$ and for $x$ of O(1) the mixing is large. Thus, in the limit of neglecting small mass
terms of order $m_{1,2}$, the demands of large atmospheric neutrino mixing and dominance of $m_3$ translate into the
condition that the subdeterminant 23 of the 3 by 3 mass matrix vanishes. The problem is to show that this
vanishing can be arranged in a natural way without fine tuning.

Without loss of generality we can go to a basis where both the charged lepton Dirac mass
matrix $m_D^l$ and the Majorana matrix $M$ for the right-handed neutrinos are diagonal. 
In fact, after diagonalization of the charged lepton Dirac mass
matrix, we still have the freedom of a change of basis for the right-handed neutrino fields, in that the right-handed
charged lepton and neutrino fields, as opposed to left-handed fields, are uncorrelated by the SU(2)$\times$U(1) gauge
symmetry. We can use this freedom to make the Majorana matrix diagonal: $M^{-1}=V^Td_MV$ with
$d_M={\rm Diag}[1/M_1,1/M_2,1/M_3]$. Then if we parametrize the matrix
$Vm_D$ by $z_{ab}$ we have:
\beq (m_\nu)_{ab}\equiv (m_D^TM^{-1}m_D)_{ab}=\sum_c \frac{z_{ca}z_{cb}}{M_c}.\label{cc}
\eeq From this expression we see that, while we can always arrange the twelve parameters $z_{ab}$ and $M_a$ to arbitrarily
fix the six independent matrix elements of $m_{\nu}$, case  A is special in that it can be approximately reproduced in two
particularly simple ways, without relying on precise cancellations among different terms: 
\begin{enumerate}
\item[$i$)] there are only two
large entries in the $z$ matrix,
$|z_{c2}|\sim |z_{c3}|$, and the three eigenvalues $M_a$ are of comparable magnitude (or, at least, with a less pronounced
hierarchy than for the $z$ matrix elements). Then, the subdeterminant 23 vanishes and one only needs the ratio
$|z_{c2}/z_{c3}|$ to be close to 1. This possibility was discussed in some of our previous papers
\cite{us,us3}; 
\item[$ii$)] one of the right-handed
neutrinos is particularly light and, in first approximation, it is only coupled to $\mu$ and $\tau$. Thus, $M_c\sim \eta$ 
(small) and $z_{c1}\sim 0$. In this case
the 23 subdeterminant vanishes, and again one only needs the ratio
$|z_{c2}/z_{c3}|$ to be close to 1. This possibility has been especially emphasized in 
refs. \cite{smirnov,zurab,hall,king}. 
\end{enumerate}

In a 2 by 2 matrix context (in the 23 sector), a typical example of mechanism $i$ is given by
a Dirac matrix $m_D$, defined by $\bar R m_D
L$, which takes the approximate form:
\beq
 m_D\propto 
\left[\matrix{0&0\cr x&1    } 
\right]~~~~~. 
\label{md00}
\eeq 
This matrix has the property that for a generic Majorana matrix $M$ one finds:
\beq 
m_{\nu}\propto 
\left[\matrix{x^2&x\cr x&1    } 
\right]~~~~~. 
\label{mn0}
\eeq 
The only condition on $M^{-1}$ is that the $1/M_3$ entry be non-zero. In particular all $M_i$ can well be of the same
order. As is clear, this mechanism is based on asymmetric Dirac matrices, with, in the case of the example, a large
left-handed mixing already present in the Dirac matrix. In our previous paper, ref. \cite{us3}, we argued that in an SU(5)
GUT left-handed mixings for leptons tend to correspond to right-handed mixings for $d$ quarks (in a basis where $u$ quarks are
diagonal). Since large right-handed mixings for quarks are not in contrast with experiment, viable GUT models 
can be constructed following this mechanism; they correctly reproduce the data on fermion masses and mixings. 

If, for some reason, one prefers symmetric or nearly so matrices, then one can use mechanism $ii$. 
For example, one could want to
preserve left--right symmetry at the GUT scale. Then, the observed smallness of left-handed mixings for quarks would
also demand small right-handed mixings. So we now assume that $m_D$ is nearly diagonal (always in the basis where
$m_D^l$ and $M$ are diagonal) with all its off-diagonal terms proportional to some small parameter $\epsilon$.
Starting from
\beq
 m_D\propto 
\left[\matrix{\epsilon^p&x\epsilon\cr x\epsilon&1    } 
\right],~~~~~ M^{-1}\propto 
\left[\matrix{r_2&0\cr 0&1    } 
\right]~~~~~, 
\label{mdM}
\eeq
where $x$ is of O(1) and $r_2\equiv M_3/M_2$, we obtain:
\beq 
m_{\nu}\propto 
\left[\matrix{\epsilon^{2p}r_2+x^2\epsilon^2&x\epsilon^{p+1}r_2+x\epsilon \cr
x\epsilon^{p+1}r_2+x\epsilon&x^2\epsilon^2 r_2+1    } 
\right]~~~~~. 
\label{mnM}
\eeq
For sufficiently small $M_2$  the terms in $r_2$ are dominant. For $p=1,2$, which we consider as typical cases, it is
sufficient that $\epsilon^2 r_2\gg 1$. Assuming that this condition is satisfied, consider first the case with $p=2$. We have
\beq 
m_{\nu}\propto 
x^2\epsilon^2 r_2\left[\matrix{\dd\frac{\epsilon^2}{x^2}&\dd\frac{\epsilon}{x}\cr 
& \cr
\dd\frac{\epsilon}{x}&1} 
\right]~~~~~. 
\label{mnp2}
\eeq
In this case the determinant is naturally vanishing (to the extent that the terms in $r_2$ are dominant), so that the
mass eigenvalues are widely split. However, the mixing is nominally small: $\sin{2\theta}$ is of O($2\epsilon/x$). It
could be numerically large enough if $1/x\sim 2$--$3$ and $\epsilon$ is of the order of the Cabibbo angle $\epsilon\sim
0.20$--$0.25$. This is what we call ``stretching'': the large neutrino mixing is explained in terms of a small parameter; this
is not so small and can give a perhaps sufficient amount of mixing if enhanced by a possibly large coefficient. This
minimalistic view was endorsed in refs. \cite{minimal}. 

A more peculiar case is obtained for $p=1$, which gives:
\beq 
m_{\nu}\propto 
\epsilon^2 r_2\left[\matrix{1&x\cr x&x^2} 
\right]~~~~~. 
\label{mnp1}
\eeq
In this case the small parameter $\epsilon$ is completely factored out and for $x\sim 1$ the mixing is nearly maximal.
The see-saw mechanism has created {\it large mixing from almost nothing}: all
relevant matrices entering the see-saw mechanism are nearly diagonal. Clearly, the crucial factorization of the small
parameter $\epsilon^2$ only arises for $p=1$, that is the light Majorana eigenvalue is coupled to $\nu_\mu$ and $\nu_\tau$
with comparable strength.

\section{Generalization to the $3\times 3$ Case}
It is straightforward to extend the previous model to the 3 by 3 case. One simple class of examples is the following one.
We start from
\beq
 m_D=v 
\left[\matrix{\epsilon''&\epsilon'& y~\epsilon'\cr
\epsilon'&\epsilon &x~\epsilon\cr y~\epsilon' &x~\epsilon &1      } 
\right],~~~~~ M^{-1}=\frac{1}{\Lambda} 
\left[\matrix{ r_1&0&0\cr 0& r_2&0\cr 0&0& r_3    } 
\right]~~~~~,
\label{mdM3}
\eeq
where, unless otherwise stated, $x$ and  $y$ are O(1); $\epsilon$, $\epsilon'$ and $\epsilon''$
are independent small numbers and $r_i\equiv M_3/M_i$. We
expect $\epsilon''\ll\epsilon'\ll\epsilon\ll 1$ and, perhaps, also $r_1\gg r_2\gg r_3=1$, 
if the hierarchy for right-handed
neutrinos follows the same pattern as for known fermions. Depending on the relative size of the ratios $r_i/r_j$,
$\epsilon/\epsilon'$ and $\epsilon'/\epsilon''$, we can have models with dominance of any of the $r_{1,2}$. 

For example, we set $x=1$ (keeping $y$ of O(1)) and
assume $r_2\epsilon^2\gg r_1\epsilon'^2,r_3$, together with  $r_2 \epsilon'^2\gg r_1 \epsilon''^2$ and 
$r_2\epsilon\gg r_1 \epsilon''$. In this limit, with good accuracy we obtain:
\beq 
m_{\nu}=
\frac{v^2}{\Lambda}r_2\epsilon^2\left[\matrix{\dd{\epsilon'^2\over \epsilon^2}&\dd{\epsilon'\over\epsilon}&
\dd{\epsilon'\over\epsilon}\cr
\dd{\epsilon'\over\epsilon}&1+\dd\frac{r_1\epsilon'^2}{r_2\epsilon^2}&1 \cr
\dd{\epsilon'\over\epsilon}&1&1+\dd\frac{r_3}{r_2\epsilon^2}    } 
\right]~~~~~. 
\label{mnFI}
\eeq 
Note that, for $\epsilon\epsilon''\le\epsilon'^2$:
\beq
{\rm Det}[m_{\nu}]=\left(\frac{v^2}{\Lambda}\right)^3 r_1 r_2 r_3 \epsilon'^4~~~~~.
\label{det}
\eeq
Approximate eigenvalues, in units of $v^2/\Lambda$ and with $m_3>m_2>m_1$, are given by
\beq
m_3\sim 2 r_2\epsilon^2,~~~~~~m_2\sim \frac{1}{2}r_1\epsilon'^2,~~~~~~m_1\sim 
\frac{r_3\epsilon'^2}{\epsilon^2}\label{la1}
\eeq
for $r_1\epsilon'^2>r_3$ or, alternatively,
\beq
m_3\sim 2 r_2\epsilon^2,~~~~~~m_2\sim \frac{1}{2}r_3,~~~~~~m_1\sim 
\frac{r_1\epsilon'^4}{\epsilon^2}\label{la2}
\eeq
for $r_1\epsilon'^2<r_3$. Having set $x=1$ the atmospheric neutrino mixing is nearly maximal. 
The solar neutrino mixing is instead generically 
small in these models, being proportional to $\epsilon'/\epsilon$. Thus the SA-MSW solution is obtained. It is easy to find
set of parameter values that lead to an acceptable phenomenology within these solutions. 
As an illustrative example we take:
\beq
\epsilon\sim \lambda^4~~,~~~~~\epsilon'\sim \lambda^6~~,~~~~~\epsilon''\sim\lambda^{12}~~,
\eeq
\beq
r_1\sim\lambda^{-12}~~,~~~~~r_2\sim\lambda^{-9}~~,
\eeq
where $\lambda\sim\sin\theta_C$, $\theta_C$ being the Cabibbo angle. The neutrino mass matrices become
\beq
m_D=v
\left[\matrix{\lambda^{12}&\lambda^6&\lambda^6\cr
\lambda^6&\lambda^4&\lambda^4\cr \lambda^6&\lambda^4&1      } 
\right]~~,~~~~~~M=\Lambda
\left[\matrix{\lambda^{12}&0&0\cr
0&\lambda^9&0\cr 0&0&1      } 
\right]
\label{ex1}
\eeq
and, in units of $v^2/\Lambda$, we obtain
\beq
m_3\sim 1/\lambda~~,~~~~~m_2\sim 1~~,~~~~~m_1\sim\lambda^4~~.
\eeq
The solar mixing angle $\theta_{12}$ is
of order $\lambda^2$, suitable to the SA-MSW solution. Also $\theta_{13}\sim\lambda^2$.

It is also possible to arrange the
parameters in eq. (\ref{mdM3}) in such a way that $r_1$ is dominant. Phenomenologically viable models can also be
constructed in this case, always of the SA-MSW type. 
For instance, we can take $y=1$ (keeping $x$ of O(1)) in order to obtain a large atmospheric mixing angle, 
and further assume the dominance of the $r_1$ terms, namely:
$\epsilon'^2 r_1\gg \epsilon^2 r_2,~r_3$, with $\epsilon''^2 r_1\gg \epsilon'^2 r_2$ and 
$\epsilon'' r_1\gg \epsilon r_2,~r_3$. 
In this case we can approximate the light neutrino
mass matrix as follows:
\beq 
m_{\nu}=
\frac{v^2}{\Lambda}r_1\epsilon'^2
\left[\matrix{\dd{\epsilon''^2\over\epsilon'^2}
&\dd{\epsilon''\over\epsilon'}&\dd{\epsilon''\over\epsilon'}\cr
\dd{\epsilon''\over\epsilon'}&1+\dd\frac{r_2\epsilon^2}{r_1\epsilon'^2}&1 \cr
\dd{\epsilon''\over\epsilon'}&1&1+\dd\frac{r_3}{r_1\epsilon'^2}    } 
\right]~~~~~. 
\label{mnFII}
\eeq 
As an example we consider the following choice of parameters:
\beq
\epsilon\sim \lambda^4~~,~~~~~\epsilon'\sim \lambda^6~~,~~~~~\epsilon''\sim\lambda^8~~,
\eeq
\beq
r_1\sim\lambda^{-14}~~,~~~~~r_2\sim\lambda^{-6}~~.
\eeq 
In this case we have
\beq
 m_D=v
\left[\matrix{\lambda^8&\lambda^6&\lambda^6\cr
\lambda^6&\lambda^4&\lambda^4\cr \lambda^6&\lambda^4&1      } 
\right]~~,~~~~~~M=\Lambda
\left[\matrix{\lambda^{14}&0&0\cr
0&\lambda^6&0\cr 0&0&1      } 
\right]~~~~~.
\label{ex2}
\eeq
In units of $v^2/\Lambda$, we find
\beq
m_3\sim 1/\lambda^2~~,~~~~~m_2\sim 1\gg m_1
\eeq
and $\theta_{12}\sim\lambda^2$.

An interesting feature of these textures, in connection with a possible realization within a GUT scheme,
is that $m_3$ is much larger than $v^2/\Lambda$ by construction. This means that the lepton number breaking scale
$\Lambda$ can be pushed up to the order of the GUT scale or even beyond. A large value of $\Lambda$, $\Lambda\geq M_{GUT}$, 
is naturally expected in SU(5) if right-handed
neutrinos are present and in SO(10) if the grand unified group is broken down to SU(5) close to or slightly
below the Planck mass. We recall that if $m_3$ is approximately given by $v^2/\Lambda$ and $v\sim 250~{\rm GeV}$,
then $\Lambda\sim 10^{15}~{\rm GeV}$.

Models based on symmetric matrices are
directly compatible with left--right symmetry and therefore are naturally linked with SO(10). This is to be confronted with
models that have large right-handed mixings for quarks, which, in SU(5), can be naturally translated into large left-handed
mixings for leptons. In this connection it is interesting to observe that the proposed textures for the neutrino Dirac
matrix can also work for up and down quarks. For example, the matrices
\beq
 m_D^u\propto
\left[\matrix{0&\lambda^6&\lambda^6\cr
\lambda^6&\lambda^4&\lambda^4\cr \lambda^6&\lambda^4&1      } 
\right]~~,~~~~~~ m_D^d\propto
\left[\matrix{0&\lambda^3&\lambda^3\cr
\lambda^3&\lambda^2&\lambda^2\cr \lambda^3&\lambda^2&1      } 
\right]~~,
\label{mud}
\eeq
where for each entry the order of magnitude is specified in terms of $\lambda\sim \sin{\theta_C}$, lead to acceptable mass
matrices and mixings. In fact $m_u:m_c:m_t=\lambda^8:\lambda^4:1$ and $m_d:m_s:m_b=\lambda^4:\lambda^2:1$. 
The $V_{CKM}$ matrix receives a
dominant contribution from the down sector in that the up sector angles are much smaller than the down sector ones. 
The same kind of texture can also be adopted in the charged lepton sector. 
It is simple to realize that models of this sort cannot be derived 
from a broken U(1)$_{\rm F}$ horizontal symmetry, with a single field that breaks spontaneously U(1)$_{\rm F}$. 
For symmetric matrices, left and right U(1) charges must be equal: then, for 
example, if $m_{33}\sim l^{2q_3} \sim 1$ or $q_3=0$, then $m_{22}\sim l^{2q_2}$ is different from 
$m_{23}\sim l^{q_2}$ etc. 

However, the symmetry requirement is not really necessary to produce a large atmospheric mixing angle from 
neutrino mass matrices characterized by small mixings. Indeed, by considering 
for simplicity the 2 by 2 case, we could equally well have started from a Dirac mass matrix of the kind:
\beq
 m_D\propto 
\left[\matrix{\epsilon&x\epsilon\cr x'&1    } 
\right]~~,~~~~~ M^{-1}\propto 
\left[\matrix{r_2&0\cr 0&1    } 
\right]~~~~~,
\label{mdMM}
\eeq
with $x'$ any number smaller than or equal to one. Once the condition $\epsilon^2 r_2\gg1$ is satisfied,
the neutrino mass matrix of eq. (\ref{mnp1}) is obtained again, independently from $x'$. 
The generalization of this mechanism to the full 3 by 3
case is straightforward, and analogous to the one discussed above for symmetric matrices.
This leaves much more freedom for model building and does not necessarily bound the realization of
these schemes to left--right-symmetric scenarios.

The solar mixing angle is generically small in the class of models explicitly discussed above. However,
small mixing angles in the Dirac and Majorana neutrino mass matrices do not exclude a large solar mixing angle.
For instance, this is generated from the asymmetric, but nearly
diagonal mass matrices:
\beq
m_D=v
\left[\matrix{\lambda^6&\lambda^6&0\cr
0&\lambda^4&\lambda^4\cr 0&0&1      } 
\right]~~,~~~~~~M=\Lambda
\left[\matrix{\lambda^{12}&0&0\cr
0&\lambda^{10}&0\cr 0&0&1      } 
\right]~~~~~.
\label{ex22}
\eeq
They give rise to a light neutrino mass matrix of the kind:
\beq 
m_\nu=
\left[
\matrix{
\lambda^2&\lambda^2&0\cr
\lambda^2&1&1\cr
0&1&1}\right]{v^2\over \lambda^2\Lambda}~~~~~,
\label{mnu3}
\eeq 
which is diagonalized by large $\theta_{12}$ and $\theta_{23}$ and small $\theta_{13}$.
The mass hierarchy is suitable to the large angle MSW solution. 

\section{Renormalization Effects}
These results are stable under renormalization from the high--energy 
scale $\Lambda$ where the mass matrices are produced down to
the electroweak scale. Indeed, suppose that we start with a 
supersymmetric theory whose superpotential $w$ at the scale 
$\Lambda\ge M_{GUT}$ is given by:
\beq
w=U^c y^u Q H_u+D^c y^d Q H_d+N^c y^\nu L H_u+E^c y^e L H_d
  +{1\over 2} N^c M N^c~~~,
\eeq
where 
\beq
y^e(\Lambda)=    
\left[
\matrix{0& 0\cr 0& y_\tau}
\right]~~,~~~~
y^\nu(\Lambda)=    
\left[
\matrix{\epsilon& x~\epsilon\cr x~\epsilon& 1}
\right]~~,~~~~
M(\Lambda)=    
\left[
\matrix{M_2& 0\cr 0& M_3}
\right]~~.
\eeq
Here we focus on the two heaviest generations. The matrices 
$y^{u,d,e,\nu}$ describe the Yukawa couplings of the theory,
related to the mass matrices through $m_{u,\nu}=y^{u,\nu} v_u$ 
and $m_{d,e}=y^{d,e} v_d$, where $v_{u,d}$ are the
vacuum expectation values of the neutral scalar components
in the superdoublets $H_{u,d}$. The mass scales
are ordered according to $\Lambda\ge M_3\gg M_2\gg M_W$,
$M_W$ denoting the electroweak scale. 
At the scale $M_2$, after integrating out the superfields
$N^c$ associated to the heavy right-handed neutrinos, the 
mass matrix for the light neutrinos, including the 
renormalization effects, is given by:
\beq
m_\nu(M_2)=
{v_u^2 \over M_2}
\left[
\matrix{({y^\nu}_{22})^2& y^\nu_{22}~ y^\nu_{23}\cr
y^\nu_{22}~ y^\nu_{23}& ({y^\nu}_{23})^2} 
\right]
+ 
O({1\over M_3})~~~.
\eeq
The couplings ${y^\nu}_{22}$ and ${y^\nu}_{23}$ are evaluated
at the scale $M_2$. If the running of $M$ is computed by
by taking $\epsilon=0$, then $M_2$ stays approximately constant when 
going from $\Lambda$ down to $M_2$. Notice that, if we neglect the terms
of $O(1/M_3)$, the determinant of $m_\nu(M_2)$ vanishes,
even including the renormalization effects. These effects 
modify the mixing angle at the scale $M_2$ and they depend on how
the Yukawa coupling ${y^\nu}_{22}$ and ${y^\nu}_{23}$ change
from $\Lambda$ to $M_2$. The running of the neutrino Yukawa couplings
is governed by the equation:
\beq
{d y^\nu\over dt}={1\over 16\pi^2} y^\nu 
\left[
{\rm tr}(3 {y^u}^\dagger y^u+{y^\nu}^\dagger y^\nu)
-4\pi(3 \alpha_2+{3\over 5}\alpha_1)
+3 {y^\nu}^\dagger y^\nu+ {y^e}^\dagger y^e
\right]~~~,
\eeq
from $\mu=\Lambda$ to $\mu=M_3$, with $t={\rm log}\mu$.
At $M_3$, the superfield $N^c_3$ is integrated out.
This produces terms of order $1/M_3$, which are neglected
in the present discussion. From $\mu=M_3$ down to
$\mu=M_2$ the running of ${y^\nu}_{22}$ and ${y^\nu}_{23}$
is described by an equation formally identical to
the previous one, with 
\beq
y^\nu\equiv
\left[
\matrix{y^\nu_{22}& y^\nu_{23}\cr y^\nu_{32}& y^\nu_{33}}
\right]
\to
y^\nu\equiv
\left[y^\nu_{22}~~~ y^\nu_{23}\right]~~.
\eeq
To assess the size of the renormalization effects on 
${y^\nu}_{22}$ and ${y^\nu}_{23}$, we have integrated numerically
the above equations in the approximation of constant $y^{u,e}$
and $\alpha_{1,2}$. This should provide a rough estimate of the effects.
As an example we take $\Lambda=10^{18}$ GeV,
$M_3=10^{16}$ GeV and $M_2=10^8$ GeV, $\alpha_2=(5/3) \alpha_1=1/24$; 
we choose $y_\tau=0.6$, corresponding
to the large $\tan\beta\equiv v_u/v_d$ regime, which enhances
the renormalization effects of $y^\nu$.
Assuming as initial conditions 
${y^\nu}_{22}(\Lambda)={y^\nu}_{23}(\Lambda)={y^\nu}_{32}(\Lambda)=0.05$ and 
${y^\nu}_{33}(\Lambda)=1$,
we find that ${y^\nu}_{22}$ and ${y^\nu}_{23}$
are modified by about 10\% at the
scale $M_2$. 
The mixing angle, nearly maximal at the scale $\Lambda$, 
remains close to maximal:
$\sin^2 2\theta_{23}\sim 0.98$ in the example at hand.

The running from $M_2$ down to the electroweak scale does
not appreciably affect neither the vanishing determinant
condition, nor a nearly maximal neutrino mixing angle.
This can be seen by analyzing the renormalization
group equation for $m_\nu$, which, in the one--loop approximation,
is given by:
\beq
{d m_\nu\over d t}={1\over 8\pi^2}
\left\{
\left[
{\rm tr}(3 {y^u}^\dagger y^u)
-4\pi(3 \alpha_2+{3\over 5}\alpha_1)
\right]m_\nu
+{1\over 2}
\left[
m_\nu {y^e}^\dagger y^e +({y^e}^\dagger y^e)^T m_\nu
\right]
\right\}~~~.
\eeq
The solution is:
\beq
m_\nu(M_W)
=
C
\left[
\matrix{{m_\nu}_{22}(M_2)&  {m_\nu}_{23}(M_2)~B\cr
{m_\nu}_{23}(M_2)~B& {m_\nu}_{33}(M_2)~B^2}
\right]~~,
\label{sol1}
\eeq
where
\beq
C={\rm exp}\left[{\dd{\int_{t_0}^t c(t') dt'}}\right]~~,~~~~
B={\rm exp}\left[{\dd{\int_{t_0}^t b(t') dt'}}\right]~~,~~~~~~~
t_0={\rm log}(M_W)~~,~~~~t={\rm log}(M_2)~~
\eeq
and
\beq
c(t)={1\over 8\pi^2}
\left[
{\rm tr}(3 {y^u}^\dagger y^u)-4\pi(3 \alpha_2+{3\over 5}\alpha_1)
\right](t)~~,
\eeq
\beq
b(t)={y_\tau^2(t)\over 16\pi^2}~~.
\eeq
We see from eq. (\ref{sol1}) that if the determinant of $m_\nu$
vanishes at the scale $M_2$, then it vanishes also at the weak
scale. The mixing angle at the weak scale is given by:
\beq
\sin^2(2 \theta)(M_W)=\frac{4 {m_\nu}_{23}^2(M_2) B^2}
{4 {m_\nu}_{23}^2(M_2) B^2+
({m_\nu}_{33}(M_2) B^2-{m_\nu}_{22}(M_2))^2}~~~.
\eeq
The condition of maximal mixing at the scale $M_2$ is
${m_\nu}_{33}(M_2)={m_\nu}_{22}(M_2)$. It is easy to see
that, if this condition is met, then the first correction to
$\sin^2(2 \theta)(M_W)$ is of second order in the parameter
$(B-1)$. Numerically, we find that $\sin^2(2 \theta)(M_W)$ is 
reduced only by a of few percent.

Summarizing, the renormalization effects do not appreciably
modify the results obtained at the tree level for eigenvalues
and mixing angle of the neutrino mass matrix. They
manifest themselves as $O(1)$ coefficients that, for all
practical purposes, can be
absorbed in the definition of the classical parameters.
An analogous analysis can be carried out in the 3 by 3 case,
with similar results. This is in agreement with the analysis
of ref. \cite{ell}, as far as the case of hierarchical neutrino
masses is concerned.

\section{An Explicit GUT Example with Broken Flavour Symmetry}
From our general discussion it is clear that some amount of correlation between the Dirac and the Majorana sectors
is required in order to reproduce, at the order-of-magnitude level, the observed
pattern of oscillations (assuming from now on the SA-MSW solution for solar neutrinos). In order to show that
this correlation could be induced by an underlying flavour symmetry and, at the same time, to present
an explicit grand unified example of the class of textures considered here, we sketch the features of an SU(5) model
with a spontaneously broken flavour symmetry based on the gauge group U(1)$_{\rm A}\times$U(1)$_{\rm B}$.
 
The three generations are described by ${\bf \Psi_{10}}^a$ and ${\bf \Psi_{\bar 5}}^a$, $(a=1,2,3)$ 
transforming as $10$ and ${\bar 5}$ of SU(5), respectively. Three more SU(5) singlets
${\bf \Psi_1}^a$ describe the right-handed neutrinos. Here we focus on the Yukawa coupling allowed by
Higgs multiplets $\varphi_5$ and $\varphi_{\bar5}$ in the $5$ and ${\bar 5}$ SU(5) representations 
and by three flavons, $\varphi_A^-$, $\varphi_B^-$ and $\varphi_B^+$, taken, to begin with, as
SU(5) singlets. The charge assignment of the various fields is summarized in table I.
We assume that the flavon fields develop the following set of VEVs:
\beq
\langle\varphi_A^-\rangle\sim\lambda~\Lambda~~,~~~~~
\langle\varphi_B^-\rangle\sim\lambda~\Lambda~~,~~~~~
\langle\varphi_B^+\rangle\sim\Lambda~~,
\eeq
where $\Lambda$ denotes the cutoff of the theory and provides the mass scale suppressing
the higher-dimensional operators invariant under the gauge and the flavour groups.
\\[0.1cm]                      
{\begin{center}
\footnotesize
\begin{tabular}{|c|c|c|c|c|c|c|c|c|}   
\hline                        
 & ${\bf\Psi_{10}}$ & ${\bf\Psi_{\bar 5}}$ & ${\bf\Psi_1}$ & $\varphi_5$ & $\varphi_{\bar 5}$ & $\varphi_A^-$ & 
$\varphi_B^-$ & $\varphi_B^+$\\
\hline
& & & & & & & & \\
SU(5) & 10 & ${\ov 5}$ & 1 & 5 & ${\ov 5}$ & 1 & 1 & 1\\ 
& & & & & & & & \\
\hline
& & & & & & & & \\
${\rm U(1)}_{\rm A}$ & (3,2,0) & (3,1,0) & (6,3,0) & 0 & 0 & $-$1 & 0 & 0\\
& & & & & & & & \\
\hline                        
& & & & & & & & \\
${\rm U(1)}_{\rm B}$ & (1,0,0) & ($-$2,$-$2,$-$1) & (0,2,0) & 0 & 0 & 0 & $-$1 & +1\\
& & & & & & & & \\
\hline
\end{tabular}                             
\end{center}}
\vspace{3mm} 
{\bf Table~I} : Quantum numbers of matter and flavon multiplets. 
\vspace{0.5cm}

In the quark sector we obtain 
\footnote{In eq. (\ref{mquark}) the entries denoted by 1 in $m_D^u$ and $m_D^d$ 
are not necessarily equal. As usual, such a notation allows 
for O(1) deviations.}
\beq m_D^u=(m_D^u)^T=
\left[
\matrix{
\lambda^8&\lambda^6&\lambda^4\cr
\lambda^6&\lambda^4&\lambda^2\cr
\lambda^4&\lambda^2&1}
\right]v_u~~,~~~~~~~ m_D^d=
\left[
\matrix{
\lambda^6&\lambda^5&\lambda^3\cr
\lambda^4&\lambda^3&\lambda\cr
\lambda^3&\lambda^2&1}
\right]v_d~~,
\label{mquark}
\eeq 
from which we get the order-of-magnitude relations:
\bea m_u:m_c:m_t & = &\lambda^8:\lambda^4:1 \nonumber\\ m_d:m_s:m_b & = &\lambda^6:\lambda^3:1
\eea and 
\beq V_{us}\sim \lambda~,~~~~~ V_{ub}\sim \lambda^3~,~~~~~ V_{cb}\sim \lambda^2~.
\eeq 
Here $v_u\equiv\langle \varphi_5 \rangle$, $v_d\equiv\langle \varphi_{\bar 5} \rangle$.
Notice that $V_{ub}$ and $V_{us}$ are dominated by the down-quark contribution.
The mass matrix for the charged leptons  is the transpose of $m_D^d$:
\beq m_D^l=(m_D^d)^T
\eeq and we find
\beq m_e:m_\mu:m_\tau  = \lambda^6:\lambda^3:1~~~~. 
\eeq 
At this level we obtain the well-known prediction $m_b=m_\tau$, together with the unsatisfactory
relation $m_d/m_s=m_e/m_\mu$ (which, however, is an acceptable order of magnitude equality).

In the neutrino sector, the Dirac and Majorana mass matrices are
given by:
\beq m_D=
\left[
\matrix{
\lambda^9&\lambda^7&\lambda^6\cr
\lambda^6&\lambda^4&\lambda^4\cr
\lambda^3&\lambda&1}
\right]v_u~~,~~~~~~~~ M=
\left[
\matrix{
\lambda^{12}&\lambda^{11}&\lambda^6\cr \lambda^{11}&\lambda^{10}&\lambda^5\cr
\lambda^6&\lambda^5&1}
\right]\Lambda~~.
\eeq
Notice that while in the model under consideration $m_D$ is asymmetric, it is diagonalized by small
unitarity transformations, i.e. transformations that go to the identity in the limit of vanishing 
$\lambda$, in both the left and the right sectors.
After diagonalization of the charged-lepton sector and after integrating out the heavy right-handed neutrinos, we obtain
the following neutrino mass matrix in the low-energy effective theory:
\beq 
m_\nu=
\left[
\matrix{
\lambda^4&\lambda^2&\lambda^2\cr
\lambda^2&1&1\cr
\lambda^2&1&1}\right]{v_u^2\over \lambda^2\Lambda}~~~~.
\label{mnu}
\eeq

The O(1) elements in the 23 block are produced only by the interplay of the left-handed mixing in $m_D$
and the hierarchy in the Majorana sector. The contribution from the charged-lepton sector is subleading.
An important property of $m_\nu$ is that, as a result of the see-saw mechanism and of the specific U(1)$_{\rm F}$ 
charge assignment, the determinant of the 23 block is of O$(\lambda^2)$ (in units of $(v_u^2/\lambda^2\Lambda)^2$).
This gives rise to the desired hierarchy between atmospheric and solar frequencies.
It is easy to verify that the eigenvalues of $m_\nu$ satisfy the relations:
\beq 
m_1:m_2:m_3  = \lambda^8:\lambda^2:1~~.
\eeq 
The atmospheric neutrino oscillations require 
$m_3^2\sim 3.5\times 10^{-3}~{\rm eV}^2$. 
From eq. (\ref{mnu}), taking $v_u\sim 250~{\rm GeV}$, the mass scale ${\Lambda}$ of the
heavy Majorana neutrinos turns out to be close to $2\times 10^{16}~{\rm GeV}$.
The squared mass difference between the lightest states is  of O$(\lambda^4)~m_3^2$,
appropriate to the MSW solution of the solar neutrino problem. Finally, beyond the large mixing in the 23 sector,
$m_\nu$  provides a mixing angle $\theta_{12} \sim \lambda^2$ in the 12 sector, close to the range preferred by the small-angle
MSW solution. In general $U_{e3}$ is non-vanishing, of O$(\lambda^2)$. We find it encouraging that the general
pattern of all that we know on fermion masses and mixings is correctly reproduced at the level of orders of magnitude. 

This model is to be taken merely as an illustration, among many other possibilities, of the scenario outlined
in this note: small mixing angles for the fundamental fermions giving rise to a large
mixing angle for the light neutrinos. For this reason we do not attempt to address some important problems,
such as the cancellation of chiral anomalies, which we implicitly postpone to an energy scale 
higher than the unification scale. We have not dealt here with the problem of recovering the correct 
vacuum structure by minimizing the effective potential of the theory.
Also, the order of magnitude description offered by this model is not intended to account for all the details of
fermion masses. Even neglecting the parameters associated with the CP violating observables, some of the relevant
observables are somewhat marginally reproduced. For instance,
a common problem of all SU(5) unified theories, based on a minimal Higgs structure, is represented by the relation
$m_D^l=(m_D^d)^T$, which, while leading to the successful $m_b=m_\tau$ boundary condition at the GUT scale, provides the
wrong prediction
$m_d/m_s=m_e/m_\mu$. We might overcome this problem
and improve the picture \cite{eg} by assuming that the flavon fields $\varphi_A^-$, $\varphi_B^{\pm}$
or a suitable subset of them transform in the adjoint representation of SU(5).
The product $(24)^q {\bar 5}$ (where $q$ is a positive integer)
contains both the ${\bar 5}$ and the $\overline{45}$ SU(5) representations and the couplings would 
allow for a differentiation between the down quarks and the charged leptons. 
Again, since the purpose of the present model was only illustrative, we do not 
insist here on recovering fully realistic mass matrices. Finally,
additional contributions to flavour-changing processes and to CP-violating observables are generally expected in a 
supersymmetric GUT. However, a reliable estimate of the corresponding effects
would require a much more detailed definition of the theory than attempted here. Crucial ingredients such as the mechanism
of supersymmetry breaking and its transmission to the observable sector have been ignored in the present note. 

In this example we have given up a possible symmetric structure of the textures in favour of
a relatively simple and economic realization of the underlying flavour symmetry.
It is possible, within the SU(5) or SO(10) gauge theories, to provide slightly more elaborate 
examples where this symmetric structure is manifest.
In the present note we have insisted on requiring both the Dirac matrices and the Majorana matrix
$M$ to be quasi-diagonal (that is with all off-diagonal entries {\it ij} suppressed with respect to the 
largest of the diagonal {\it ii} and {\it jj} ones). 
In some alternative models, all Dirac matrices are indeed quasi-diagonal, but large mixings are
present in the Majorana matrix $M$. For instance,
the idea of a large atmospheric mixing angle from intrinsically
small mixings in the Dirac mass matrices has recently been advocated
in the context of a $U(2)$ flavour theory \cite{bcr}. In this model the Dirac neutrino mass matrix is  
the one displayed in eq. (\ref{mdM3}), with $y$ of O$(\epsilon)$ (i.e. $(m_D)_{13}\ll (m_D)_{12}$) 
and $\epsilon''\sim\epsilon'^2$. The $U(2)$ symmetry favours an $r_1$ dominance, by adequately
suppressing the $M_{11}$ element. However, to have both $\nu_\mu$ and $\nu_\tau$
with approximately the same coupling to the light Majorana mass eigenstate,
this should possess a component of order $\epsilon'$ along the $\nu^c_3$
direction. In turn, this property is only exhibited if also the $M_{33}$
entry is sufficiently small
\footnote{Technically, this suppression can be realized in the framework of 
an $SO(10)$ gauge group, by assuming that there are no flavour singlets
that couple to $\nu^c_3\nu^c_3$.}. As a consequence, there are two heavy Majorana
states with similar masses and there is a large mixing in the 23 Majorana
sector. In this case, large mixings, which are absent from the Dirac mass matrices, arise from the Majorana sector, 
a physically similar possibility but technically different from that examined here.

\section{Conclusions}
Summarizing, in most models that describe neutrino oscillations with nearly maximal mixings, there appear
large mixings in at least one of the matrices $m_D$, $m_D^l$, $M$ (i.e. the neutrino and charged-lepton
Dirac matrices and the right-handed Majorana matrix). In this note we have discussed the peculiar possibility 
that large neutrino mixing is only produced by the see-saw mechanism starting from all nearly diagonal matrices.
Although this possibility is certainly rather special, we have shown that models of this sort can
be constructed without an unrealistic amount of fine tuning and are well compatible with grand unification ideas
and the related phenomenology for quark and lepton masses.

\noindent
{\bf Acknowledgements}
One of us (F.F.) would like to thank the CERN Theory Division for its kind hospitality during July,
when this work was completed. He is also grateful to Jean-Pierre Hurni and Anna Rossi for useful discussions.

\end{document}